\documentclass[conference, 10pt,onecolumn]{IEEEtran}

\usepackage{cite}
\usepackage[cmex10]{amsmath}
\usepackage{graphicx}
\usepackage[caption=false,font=footnotesize]{subfig}
\usepackage{amssymb}
\usepackage{multirow}
\usepackage{algorithmic}
\usepackage{algorithm}

\DeclareMathOperator{\maximize}{maximize}

\allowdisplaybreaks

\begin{document}

\title{Joint User Association and Reuse Pattern Selection in Heterogeneous Networks}

\author{\IEEEauthorblockN{Quan Kuang}
\IEEEauthorblockA{Associate Institute for Signal Processing, Technische Universit\"at M\"unchen, 80290 Munich, Germany \\
Email: quan.kuang@tum.de}}



\maketitle

\begin{abstract}

The successful deployment of LTE heterogeneous networks (HetNets) depends crucially on the inter-cell interference (ICI) management. Among ICI coordination schemes, fractional frequency reuse (FFR) is considered as an efficient technique well-suited to OFDMA-based HetNets. Two coupled questions in this context are: 1) how to associate users to appropriate base-stations considering the long list of available candidate cells, and 2) how to allocate frequency resources among multiple cells. In this paper, we treat the multi-cell frequency allocation as frequency partitioning among multiple \emph{reuse patterns}, and develop a novel algorithm to solve these two coupled questions in a joint manner. We also provide practical criterion to select the set of essential candidate patterns from all possible patterns. Results show that the proposed joint strategy improves both the cell-edge user and overall network throughput.

\end{abstract}

\section{Introduction}


The heterogeneous network (HetNet), where low-power low-complexity base-stations (BSs) are overlaid with conventional macro BSs, is being considered as a promising paradigm for increasing system capacity and coverage in a cost-effective way. Due to Orthogonal Frequency Division Multiple Access (OFDMA) mechanism adopted in these networks, intra-cell interference is nearly null. However, the inter-cell interference (ICI) potentially introduced by hierarchical layering of cells becomes a fundamental limiting factor to the HetNet performance. To cope with ICI, fractional frequency reuse (FFR), which was originally proposed in conventional macro-only networks \cite{Boudreau2009}, is considered to be an efficient technique for OFDMA-based HetNets \cite{Saquib2013}. Compared to other ICI coordination schemes, the FFR method requires minimal cooperation among BSs and has a less complex operational mechanism, attracting the increasing research efforts \cite{Saquib2013}.

\begin{figure}[!t]
\centering
\includegraphics[width=3.5In]{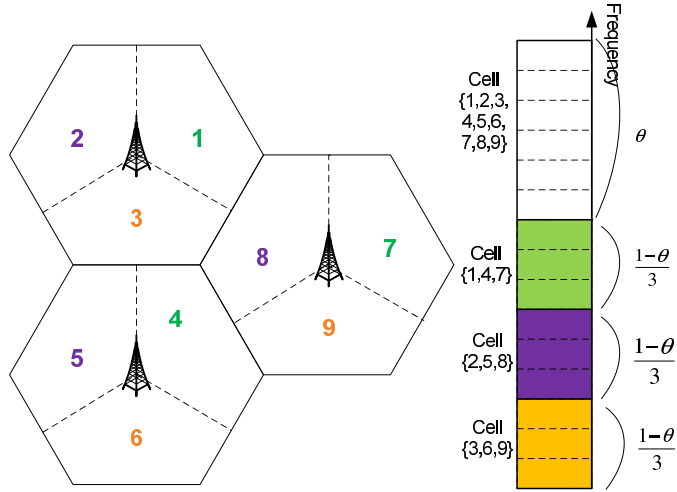}
\caption{An illustration of fractional frequency reuse by combining reuse-1 and reuse-3 patterns in a macro-only network. Reuse-1 pattern activate cells \{1,2,3,4,5,6,7,8,9\}, whereas three reuse-3 patterns activate \{1,4,7\}, \{2,5,8\}, and \{3,6,9\}, respectively.}\label{fig_macroFFR}
\end{figure}

In a regular macro-only network as shown in Fig.~\ref{fig_macroFFR}, the design of FFR is relatively a simple task. The basic mechanism is to share the available frequency band among a reuse-1 pattern and three reuse-3 patterns. The \emph{reuse pattern} (or simply \emph{pattern}) is defined as a combination of ON/OFF activities of BSs. For example, the reuse-1 pattern simply activates all the BSs, and a reuse-3 pattern activates only one BS among neighboring three BSs. To allocate the frequency band to different cells can be viewed equivalently as allocating frequency resources among different patterns. As a simple approach suitable in this homogeneous network, we can optimize a single parameter $\theta$ to control the bandwidth allocated to universally reused subbands, and then divided the rest of bandwidth evenly among three reuse-3 modes. As a result, each cell can schedule less vulnerable users (e.g., at cell-center) to the universal subbands and cell-edge users to the subbands with lower number of active BSs.


However, in a network consisting of macro and densely deployed small cells, the frequency reuse planning is no longer a simple task, due to the irregular cell location and overlaid cell deployment. The important challenges are:

1. How can we select the most important patterns?

2. How can we allocate the frequency resources among the selected patterns?

Obviously, the above questions are highly coupled with the user traffic distribution and user association policy. A cell with less number of associated users will likely be allocated less frequency resources. In this paper, we target at jointly optimizing the user association and multi-cell frequency allocation, and shed light on the above questions.

\subsection{Related work and contributions}

FFR has attracted lots of research efforts from both academia \cite{Chang2009, Novlan2010, Dotzler2010, Ali2009} and industrial standardizations \cite{R1-0513412005}. The joint power and frequency allocation for OFDMA FFR has also been studied, for example in \cite{Lopez-Perez2012b} and references cited therein. However, despite their coupled nature, the joint user association and frequency allocation for FFR has rarely been investigated. In \cite{Son2009}, the authors studied a dynamic user association problem in a FFR network. However, the frequency partitioning is assumed given beforehand and fixed. In \cite{Kuang2012}, the authors jointly optimized the user association, power and frequency allocation. However, the formulation and methodology only apply to the uplink sum rate maximization.

In this paper, we formulate the joint user association and multi-cell frequency allocation problem by treating the frequency allocation as frequency partitioning among multiple patterns. To solve the obtained mixed-integer nonconvex problem, we develop a tabu search algorithm. Tabu search is regarded as a metaheuristic that guides a local search procedure to explore the solution space beyond local optimality \cite{Glover1990}. Our development is based on the effective exploitation of problem structure, hence results in a efficient algorithm. The proposed scheme significantly improves cell-edge and median user throughput compared to no frequency allocation, without sacrificing the overall network throughput. We also provide practical criterion to narrow down the number of candidate patterns to reduce the complexity.

Note that reuse pattern selection has been studied in a \emph{Time Division Multiple Access (TDMA)} macro-only network in \cite{Son2011}, which is different from our formulation. Besides, the HetNet deployment requires new criterion to select essential patterns.

\section{System model}

We consider a downlink OFDMA HetNet, where a number of small cells are embedded in the conventional macro cellular network\footnote{Cell and BS are used interchangeably in this paper.}.  The set of all cells is denoted as $\mathcal{B}$, and the cardinality $B = |\mathcal{B}|$. A set of users $\mathcal{K}$, with $K = |\mathcal{K}|$,  are distributed in the network and we assume that each user is associated with only one serving cell.
Denote by $\boldsymbol\alpha = (\alpha_{kb}: k \in \mathcal{K}, b \in \mathcal{B})$ the association indicator matrix, i.e., $\alpha_{kb}=1$ if user $k$ is associated with cell $b$, and $0$ otherwise. Since each user is served by only one cell, we should have $\sum_{b \in \mathcal{B}} \alpha_{kb} = 1, \  \forall k \in \mathcal{K}$.



We use $g_{bk,n} \triangleq \sqrt{G_{bk}}h_{bk,n}$ to denote the channel gain between BS $b$ and user $k$ at subcarrier $n$, where $G_{bk}$ is the large-scale channel gain including path loss and shadowing, $h_{bk,n}$ accounts for the small-scale fading. We assume $\{h_{bk,n}, \forall b, \forall k, \forall n\}$ are independent and identically distributed (i.i.d.). Following the Shannon's formula, the ergodic rate of user $k$ served by $b$-th BS under pattern $i$, in bit/s/Hz, can be written as
\begin{equation}\label{RatePerCarrier}
  \bar{r}_{kb}^i = \mathbb{E}_{\textbf{h}_{\cdot k,1}} \left[\log_2 \left(1+ \frac{P_b^i G_{bk} \|h_{bk,1}\|^2}{\sigma^2 + \sum_{l\neq b} P_l^i G_{lk} \|h_{lk,1}\|^2}\right) \right]
\end{equation}
where $\textbf{h}_{\cdot k,1} \triangleq (h_{1k,1},h_{2k,1},\ldots, h_{Bk,1})$ with $h_{bk,1}$ denoting the fast fading coefficient between BS $b$ and user $k$ over an arbitrary frequency subcarrier, $P_b^i$ and $\sigma^2$ are the transmit power of BS $b$ per Hz and the received noise power per Hz, respectively. We assume if BS $b$ is active transmit power is evenly distribute across all frequencies, i.e., $P_b^i = P_b^\text{tx} / W $ where $P_b^\text{tx}$ is the total transmit power of BS $b$ and $W$ is the system bandwidth; otherwise $P_b^i = 0$ if BS $b$ is muted under pattern $i$. We further assume uniformly distributed noise effect over frequencies. Hence the ergodic rate of (\ref{RatePerCarrier}) is uniform across the whole bandwidth for any pattern.

Assuming that the set of users associated with the same BS share all the frequency resources according to a chosen scheduler, e.g., round-robin scheduler\footnote{The framework developed in this paper can be directly applied to proportional fairness scheduler by introducing a scheduling gain in (\ref{PatternUserRate}). Please see Eq.(8) in \cite{Son2009}}, we can denote the average rate of user $k$ under pattern $i$ as
\begin{equation}\label{PatternUserRate}
  R_k^i = \sum_{b \in \mathcal{B}} \alpha_{kb} \left(\frac{W \bar{r}_{kb}^i}{\sum_{j \in \mathcal{K}} \alpha_{jb}}\right) \ \ \text{[bit/s]}
\end{equation}
where $W$ is the total bandwidth, $\sum_j \alpha_{jb}$ represents the number of users associated with BS $b$.

Given the set of all possible reuse patterns $\mathcal{I}$, $I = |\mathcal{I}|$, $\{\bar{r}_{kb}^i\}$ can can be pre-calculated using (\ref{RatePerCarrier}) and treated as constants. The resource allocation strategy is to distribute total $W$ bandwidth among different patterns. Let $\boldsymbol{\pi} \triangleq (\pi_1, \ldots, \pi_i, \ldots, \pi_I) \in \Pi$ be the allocation profile, where $\pi_i$ represents the fraction allocated to pattern $i$ and $\Pi = \{\boldsymbol{\pi} | \sum_i \pi_i = 1, \pi_i \geq 0, \forall i\}$. Then we can express the average user rate after frequency allocation as
\begin{equation}\label{RateCombined}
  \bar{R}_k = \sum_{i \in \mathcal{I}} \pi_i R_k^i = \sum_{i \in \mathcal{I}} \pi_i \sum_{b \in \mathcal{B}} \alpha_{kb} \left(\frac{W \bar{r}_{kb}^i}{\sum_{j \in \mathcal{K}} \alpha_{jb}}\right).
\end{equation}

\section{User association and multi-cell frequency allocation}

\subsection{Problem statement and approach}

Our objective is to maximize the long-term network utility by optimizing the joint pattern selection and user association, as
\begin{IEEEeqnarray}{rCl}\label{problem1}
    \displaystyle\mathop{\maximize}_{\boldsymbol{\alpha},\boldsymbol{\pi}}
                \quad && U(\boldsymbol{\alpha},\boldsymbol{\pi}) = \sum_{k \in \mathcal{K}} \ \omega_k \log(\bar{R}_k) \IEEEyessubnumber \label{obj} \\
    \text{subject to} \quad && \alpha_{kb} \in \{0,1\}    \IEEEyessubnumber\label{integer constraint} \\
                      && \sum_{b \in \mathcal{B}} \alpha_{kb} = 1, \forall k \in \mathcal{K} \IEEEyessubnumber\\
                      && \pi_i \geq 0, \forall i \in \mathcal{I}  \IEEEyessubnumber\\
                      && \sum_{i \in \mathcal{I}} \pi_i = 1 \IEEEyessubnumber
\end{IEEEeqnarray}
where $\bar{R}_k$ is expressed in (\ref{RateCombined}), and $\omega_k$ provide a means for service differentiation. Note that our developed algorithm can also work for any other concave utility functions.

The formulation of (\ref{problem1}) is a mixed-integer nonconvex problem. Even we relax the integer constraint of (\ref{integer constraint}) to allow any continuous value between 0 and 1 for $\alpha_{kb}$, the resulting problem is still nonconvex because the objective function is not jointly concave in $\boldsymbol{\alpha}$ and $\boldsymbol{\pi}$. Thus, the globally optimal solution is prohibitive for reasonably sized network. Our contribution is to develop efficient algorithms to find out good solutions. The approach we adopted is based on the tabu search.
 Tabu search has obtained optimal and near optimal solutions to a wide variety of combinatorial problems
 \cite{Glover1990}.  In the following, we describe the details of the tabu search procedure to solve problem (\ref{problem1}).

\subsection{Tabu search procedure}

Generally speaking, tabu search starts from an initial solution and moves at each iteration from the current solution to the best one in its neighborhood, even if this leads to a deterioration of the objective function value, so as to allow escaping from local optimum. The search is guided by two types of memory: short-term and long-term memory. At each iteration, the move being performed is recorded in the short-term memory and the reverse move is forbidden (i.e., \emph{tabu}) for a certain number of iterations (\emph{tabu tenure}), to avoid cycling. The tabu status of a move can be revoked through an \emph{aspiration} criterion if this move improves the best solution found so far. The long-term memory is used to force the search into previously unexplored areas of the search space by a \emph{diversification strategy}. The key elements of our tabu search is the follows.

\subsubsection{Search space}

The definition of the search space is simply the space of all possible solutions that can be visited during the search. For the problem of (\ref{problem1}), the search space could naturally be the set of all feasible $\boldsymbol{\alpha}$ and $\boldsymbol{\pi}$.

\subsubsection{Neighborhood specification}

The basic iterative step of any tabu search procedure involves moving from a current solution in the search space to one of its "neighbors" according to suitably defined neighborhood structure. In accordance with our definition of the search space, we define the neighborhood structure by considering moves in which a single strategy is modified, i.e., the association of a single user, or the pattern profile $\boldsymbol{\pi}$ is changed. Hence, the number of neighbors in our definition is $K\times (B-1) +1$.
 One advantage of our definition of the neighborhood is that it allows efficient evaluation of the objective function for \emph{all} possible neighbors of the current solution. Note that if the integer variable $\boldsymbol{\alpha}$ is given the problem of (\ref{problem1}) becomes a convex problem in $\boldsymbol{\pi}$, which can be easily solved by some off-the-shelf approaches \cite{Boyd2004}. On the other hand, for fixed $\boldsymbol{\pi}$ the evaluation of the objective function by allowing only a single user to change its association can be directly calculated by simple arithmetic operations.

\subsubsection{Tabu list and aspiration criteria}

At current iteration, if user $k$ changes its association from BS $l$ to $m$, we declare tabu moving user $k$ back to BS $l$ no matter from which BS, and record this in the tabu list as $(k, l)$. On the other hand, if pattern profile $\boldsymbol{\pi}$ is modified at current iteration, we record $(K+1, 0)$ in the tabu list (by treating the profile parameter $\boldsymbol{\pi}$ as the $(K+1)$-th user). A new move is forbidden as long as it remains on the tabu list. The list has the short-term memory length of $r$ and is maintained on the first-in-first-out basis.

Although central to the tabu search method, tabus are sometimes too powerful. They may prohibit promising moves, or lead to an overall stagnation of the search process. Thus, a certain mechanism called aspiration criteria is used to allow cancelling tabus. The aspiration criterion used in our implementation is a simple one: a tabu move is allowed when it results in a solution with an objective value better than that of the current best-know solution.

\subsubsection{Initial solution}

Typically in cellular networks, a user equipment (UE) selects a cell for association with maximum received signal strength. In a HetNet, however, this association criterion will lead to the case where the macro BSs become resource constrained, whereas the small BSs serve very few users, due to the much stronger transmit power of macro BSs. To overcome this, LTE standards have proposed a concept called range expansion (RE) to off-load macro UEs to small cells by adding a positive bias to the downlink signal strength of small cells during the cell selection. We adopt this criterion for the determination of initial value of $\boldsymbol{\alpha}$ as
\begin{equation}\label{alphaInit}
    [\boldsymbol{\alpha}]_{k,m}=\left\{
                                        \begin{array}{rl}
                                        1 & \text{if $m = m_k^\star$} \\
                                        0 & \text{otherwise}
                                        \end{array} \right.
\end{equation}
where
\begin{equation}\label{re}
m_k^\star = \arg \max_{b \in \mathcal{B}} (P^\text{rx}_{kb} + \beta_b), \forall k
\end{equation}
where $P^\text{rx}_{kb} = P_b^\text{tx} G_{bk}$ is the received power (in dBm) of user $k$ from BS $b$,  $\beta_b$ is the bias (in dB).

Once the initial value of $\boldsymbol{\alpha}$ is determined, the initial value of $\boldsymbol{\pi}$ can be obtained by solving the convex optimization problem of (\ref{problem1}) for the given $\boldsymbol{\alpha}$.

\subsubsection{Diversification and multistart strategies}

In an early implementation of the algorithm, we found out it converges to local optima very soon (after 20-30 iterations) and gets trapped in those points, in the sense that no single strategy modification can improve the current solution. Hence, we implement a diversification step that will be executed after a certain number of iterations without improvement.

One simple diversification strategy is to re-configure cell bias value to create a new starting solution, and run the whole search process again. However, this strategy is not very efficient because improper initialization will cause significantly longer convergence time. Moreover, the knowledge of the search history is not well exploited.

So the diversification step we adopted is always performed starting from the best global solution found up to this point in the search. Based on the long-term memory, we select $\gamma$ (referred to as diversification amplitude) users with lowest activity counts (number of association modifications) from the beginning of the search process, and randomly assign new BSs for these users. Thus the new solution differs from the best global solution in exactly $\gamma$ components. Then the search is resumed from the new solution. To prevent too quick a reversal of these associations, they are added to the short-term memory.

\subsubsection{Stopping criteria}

The whole search alternates between two component methods: 1) a inner loop improves current solution by single strategy move and 2) a diversification step to create new starting solution. The inner loop is terminated after $MaxIterInner$ consecutive iterations without an improvement in the objective function. The whole search process stops after the total number of iterations reach $MaxIterTotal$.

The main steps of the searches procedure is summarized in Algorithm 1.

\begin{algorithm}
\label{alg1x}
\caption{Tabu search for joint user association and multi-cell frequency allocation}
\begin{algorithmic}[1]
\STATE \textbf{\emph{Initialize}}: $t=0$, $\mathbf{s}^0 = (\boldsymbol{\alpha}^0, \boldsymbol{\pi}^0)$, $U^0 = U(\mathbf{s}^0)$;
\STATE $\mathbf{s}_\text{opt} = \mathbf{s}^0$,
$U_\text{opt} = U^0$;
\STATE Initialize the long-term memory as empty;
\WHILE{$t \leq MaxIterTotal$}
\STATE Tabu list = $\varnothing$, $j = 0$;

\WHILE{$j \leq MaxIterInner$}
\STATE \textbf{\emph{Neighborhood formulation}}: create all possible neighbors of the current solution $\mathbf{s}^t$ and store them in $\mathcal{N}(\mathbf{s}^t)$;
\STATE \textbf{\emph{Neighbor selection}}: $\hat{\mathbf{s}} = \arg \max_{\mathbf{s} \in \mathcal{N}(\mathbf{s}^t)} U(\mathbf{s})$, and denote the move as $(k, l)$ if user $k$ association is changed to BS $l$ or as $(K+1, 0)$ if pattern profile is changed to obtain $\hat{\mathbf{s}}$;
\IF{(the move is in tabu list) AND $U(\hat{\mathbf{s}})  \leq U_\text{opt}$}
\STATE $\mathcal{N}(\mathbf{s}^t) = \mathcal{N}(\mathbf{s}^t) - \hat{\mathbf{s}}$ ;
\STATE Goes to step 8;
\ENDIF

\STATE $\mathbf{s}^{t+1} = \hat{\mathbf{s}}$; \COMMENT{Best admissible neighbor has been found}

\IF{$U(\mathbf{s}^{t+1}) \leq U(\mathbf{s}^t)$}
\STATE $j = j+1$; \COMMENT{Counter for no improvement}
\ELSIF{$U(\mathbf{s}^{t+1}) > U_\text{opt}$}
\STATE \textbf{\emph{Global solution update}}: $\mathbf{s}_\text{opt} = \mathbf{s}^{t+1}$,
$U_\text{opt} = U(\mathbf{s}^{t+1})$;
\ENDIF

\STATE \textbf{\emph{Tabu list update}}: record tabu for the current move, and delete oldest entry if necessary;
\STATE \textbf{\emph{Long-term memory update}}: record current move to the memory;
\STATE $t = t+1$;
\ENDWHILE

\STATE \textbf{\emph{diversification}}: set $\mathbf{s}^{t} = \mathbf{s}_\text{opt}$, and then modify $\mathbf{s}^{t}$ by selecting $\gamma$ users with lowest activity counts up to now and randomly assigning new BSs for these users;
\ENDWHILE
\STATE \emph{\textbf{Output}}: $\mathbf{s}_\text{opt}$.
\end{algorithmic}
\end{algorithm}

\section{Performance evaluation and discussions}
\subsection{Simulation scenarios and parameter setting}

\begin{figure}[t]
\centering
\includegraphics[width=4In]{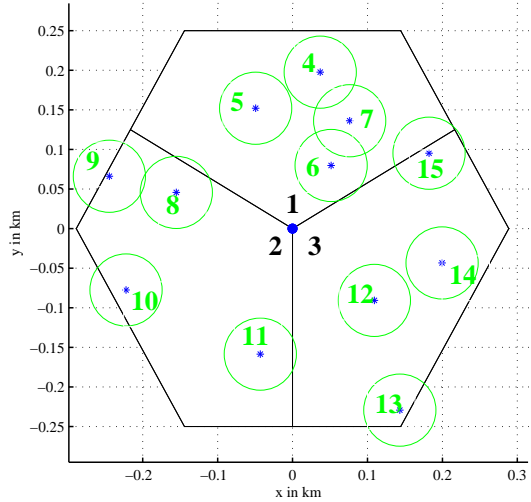}
\caption{A heterogeneous network consisting of 15 cells. }\label{fig_network}
\end{figure}

\begin{table}[t]
\renewcommand{\arraystretch}{1.0}
\caption{Network parameters.}
\label{table1}
\centering
\begin{tabular}{c c}
\hline\hline 
Parameter & Description \\ [0.5ex] 
\hline 
bandwidth & 10 MHz  \\ 
Macro total Tx power   & 46 dBm  \\
Pico total Tx power    & 30 dBm \\
Macro antenna gain & 15 dB\\
Pico antenna gain & 5 dB \\
Macro path loss & $128.1+37.6\log_{10}(R)$ \\
Pico path loss & $140.7+36.7 \log_{10}(R)$ \\
Penetration loss & 20 dB \\
Shadowing std. dev. & 8dB(macro), 10dB(pico) \\
Shadowing corr. distance & 25 m \\
Macrocell shadowing corr. & 1 between cells \\
Picocell shadowing corr. & 0.5 between cells \\
Fading model & No fast fading \\
Min. macro(pico)-UE dist. & 35 m (10 m) \\
Min. macro(pico)-pico dist. & 75 m (40 m)\\
Noise density and noise figure & -174 dBm/Hz, 9dB\\
 \hline\hline 
\end{tabular}
\end{table}

We consider a network consisting of 3 macro cells, each of which contains 4 randomly dropped pico cells as shown in Fig.\ref{fig_network}. The cells are labelled as
\begin{equation*}
  \underbrace{1,2,3}_{\text{macro cells}}, \underbrace{4,5,6,7}_\text{picos in cell 1}, \underbrace{8,9,10,11}_\text{picos in cell 2}, \underbrace{12,13,14,15}_\text{picos in cell 3}
\end{equation*}

The parameters for propagation modelling and simulations follow the suggestions in 3GPP evaluation methodology \cite{3GPP2010}, and summarized in Table \ref{table1}. We consider user density ranging from 30, 60, to 100 UEs/macro-cell. All UEs have unit weights ($\omega_k = 1$ in (\ref{obj})). In simulation, the total number of users are randomly dropped in the network, and we average over 5 drops of users and pico locations for each user case.

After tuning the tabu search parameters, we find the following setting most effective: Tabu Tenure $r= 2$ and the maximum number of non-improvement for inner loop $MaxIterInner = 4$. Using such short Tabu Tenure seems surprising at first glance, but it allows the algorithm to balance the local search and diversification, which has also been suggested in \cite{Crainic1993}. The total number of iterations $MaxIterTotal = 800$. The diversification amplitude $\gamma$ is set to 15.

\subsection{Comparison to reuse-1 baseline}

\begin{figure}[!t]
\centering
\subfloat[5th percentile of UE throughput ]{\includegraphics[width=2.5In]{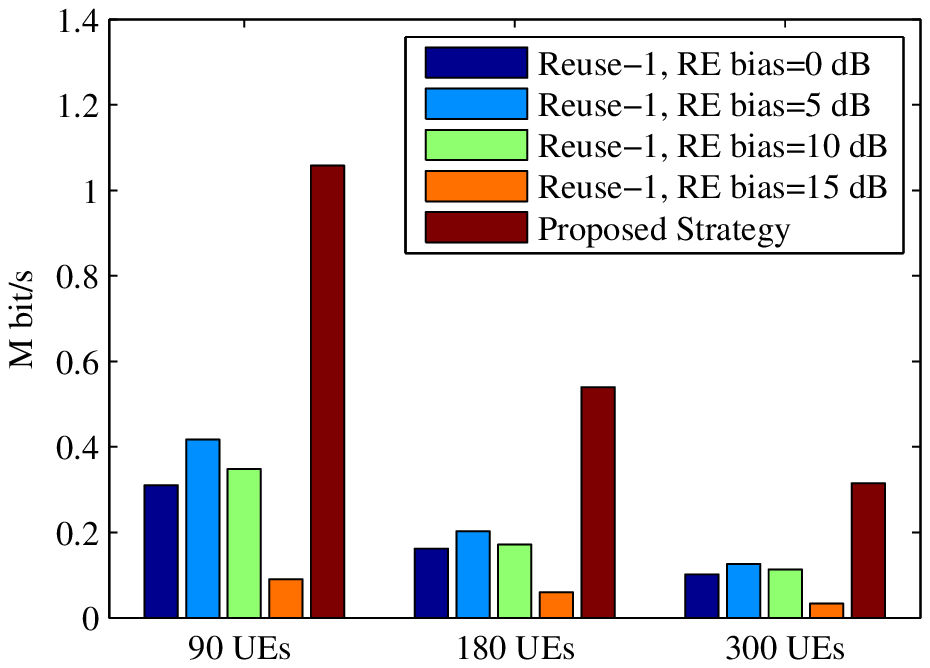}\label{fig_comp_5pct}}
\hspace{0.1pt}
\subfloat[50th percentile of UE throughput]{\includegraphics[width=2.5In]{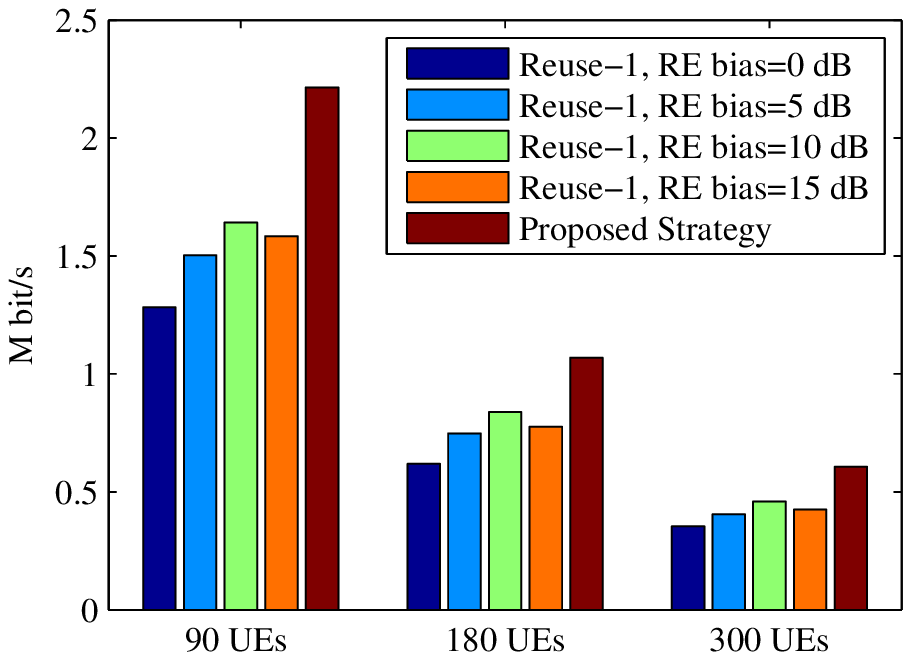}\label{fig_comp_50pct}}
\\
\subfloat[95th percentile of UE throughput]{\includegraphics[width=2.5In]{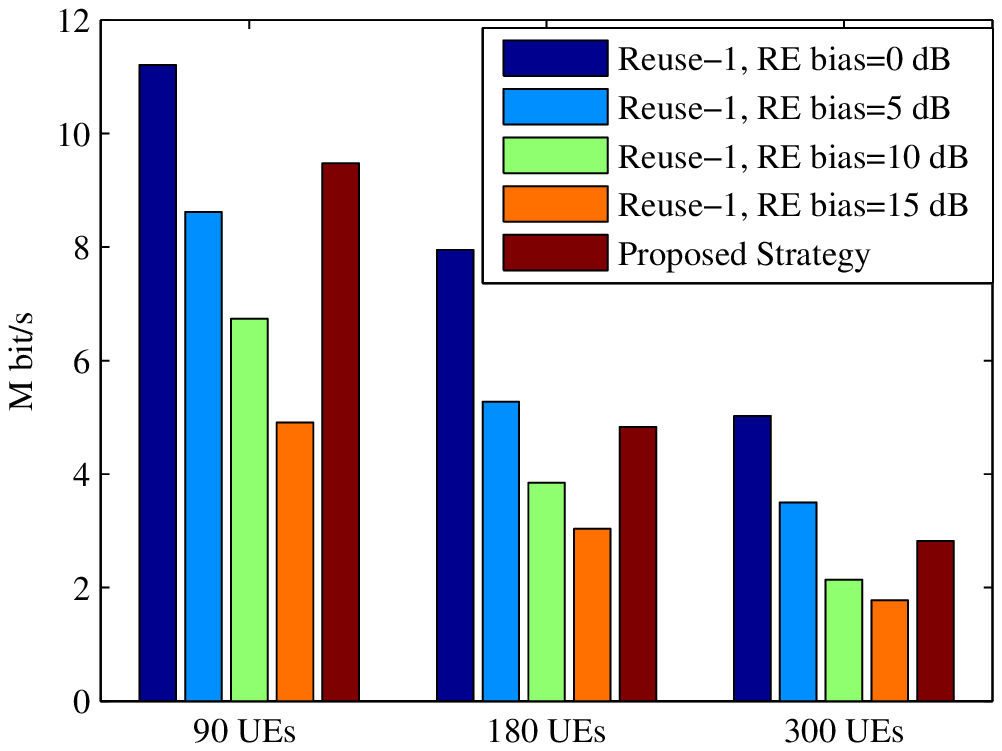}\label{fig_comp_95pct}}
\hspace{3pt}
\subfloat[Total sum of UE throughput]{\includegraphics[width=2.5In]{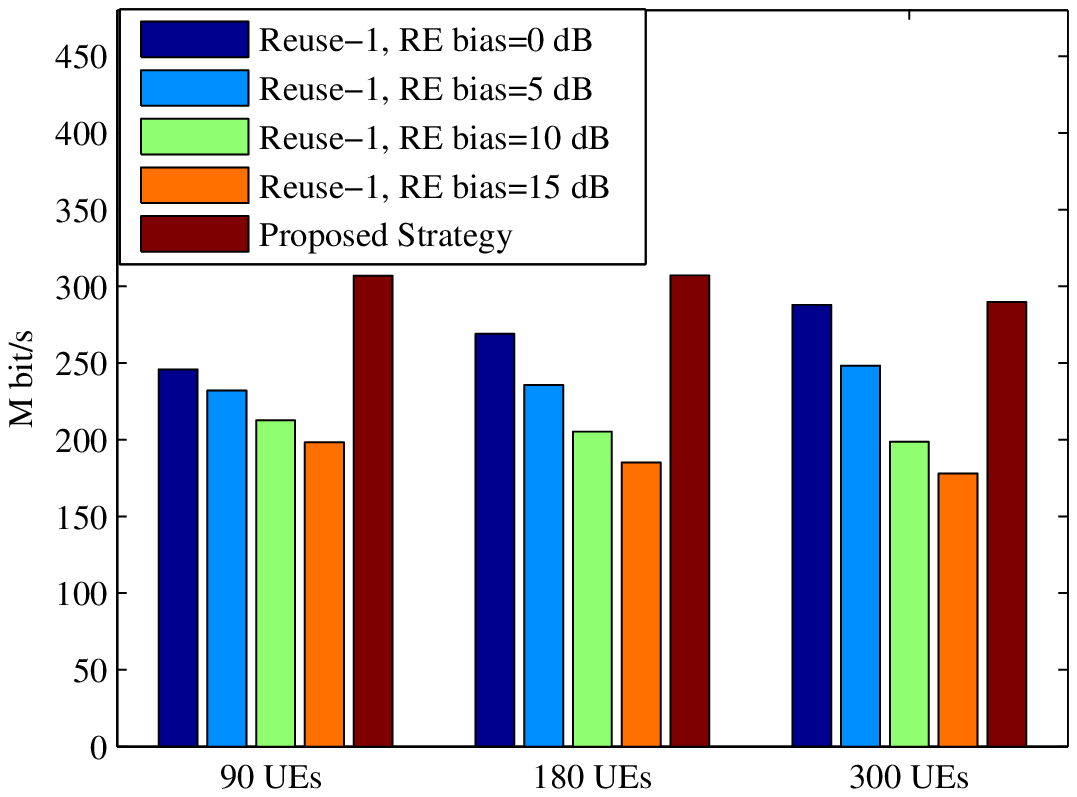}\label{fig_comp_sum}}
\caption{Comparison of proposed joint user association and frequency allocation strategy with range-expansion-based association with reuse-1 strategies. }
\label{fig_allcomp}
\end{figure}

We compare the proposed joint user association and frequency allocation strategy with the baseline system where user association is based on received power plus cell bias and reuse-1 pattern is the only pattern used. To increase the pico footprint, in the baseline system we set the macro bias to zero, and pico bias is chosen from 0, 5, 10 and 15 dB. All pico cells have the same bias value. In Fig.~\ref{fig_allcomp}, we report the UE throughput versus the total number of UEs in the network. One observation is that the 5th, 10th and 95th percentiles of UE throughput all decrease as the number of UEs increases. This is because less resources are available for individual users as we increase the total number of users while the total resources are fixed in the network. As also shown, our joint strategy significantly improves the cell-edge UEs(5th percentile throughput) and median UE throughput. Remarkably, this cell-edge improvement does not compromise the network throughput as evident in Fig.~\ref{fig_comp_sum}. Only the very-center UEs can not benefit from the interference management by the proposed strategy as shown in Fig.~\ref{fig_comp_95pct}, where reuse-1 with zero bias association gives the best performance.

Table~\ref{table2} shows that the overall log-utility of the system is better with our proposed strategy compared to the reuse-1 with different association bias values.

\begin{table}[!ht]
\renewcommand{\arraystretch}{1.0}
\caption{Comparison of total log-utility in bit/s for different strategies.}
\label{table2}
\centering
\begin{tabular}{|c | c | c | c |c | c| }
\hline
\begin{tabular}[x]{@{}c@{}}Number \\ of UEs\end{tabular} & \begin{tabular}[x]{@{}c@{}}Reuse-1, \\ 0 dB \end{tabular}  & \begin{tabular}[x]{@{}c@{}}Reuse-1, \\ 5 dB \end{tabular} & \begin{tabular}[x]{@{}c@{}}Reuse-1, \\ 10 dB \end{tabular} & \begin{tabular}[x]{@{}c@{}}Reuse-1, \\ 15 dB \end{tabular} & \begin{tabular}[x]{@{}c@{}}Proposed \\ Strategy\end{tabular} \\
\hline
90 &   1271.9 & 1284.7 & 1282.7 & 1253.7 & 1328.8 \\
\hline
180&  2412.9 &  2441.3 & 2435.7 & 2376.2 & 2535.1 \\
\hline
300 & 3854.1& 3896.8& 3896.4& 3789.0& 4054.2 \\

\hline 
\end{tabular}
\end{table}


\subsection{Feature pattern identification}

As the number of all possible pattern in the network grows exponentially with the number cells, it is necessary to pre-define a set of candidate patterns in order to reduce the complexity of the algorithm. For a macro-only homogeneous network, \cite{Son2011} suggested a guideline for candidate pattern selection in the TDMA macro networks by choosing only two kinds of patterns: 1) reuse-1: all BSs are active and 2) all neighboring BSs except one are active. However, this guideline cannot be applied to the HetNet we are investigating, because the neighboring cells are not well-defined due to the overlaid deployment. Moreover, the significantly larger number of small cells also makes this approach produce too many candidate patterns.

\begin{figure}[!t]
\centering
\includegraphics[width=3.5In]{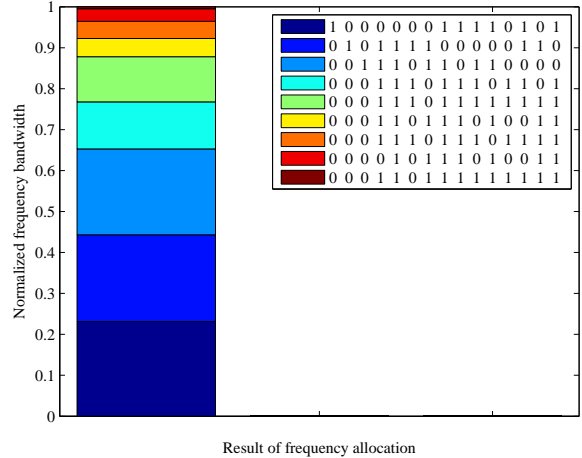}
\caption{Result of frequency allocation for a random drop of 90 UEs.}\label{fig_feature patterns}
\end{figure}

To study the feature pattern selection in HetNet, we first visualize the results of our proposed resource allocation strategy by considering \emph{all} possible patterns in the tested network. One instance is given in Fig.~\ref{fig_feature patterns}. As seen, there are only 9 out of $(2^{15} -1)$ patterns are used as the result of optimization. We use 0 at position $i$ of a row vector to indicate cell $i$ is OFF, and 1 if cell $i$ is ON. Unlike the guideline previously suggested in macro network, reuse-1 pattern actually is not included in the final result, which we also observe for other UE and pico drops. After examining all the results we tested, we propose the following practical criterion for candidate pattern selection in HetNet:

1). All macros are OFF, and all picos are ON.

2). One macro is ON among the adjacent three macros, and all picos are ON except those in the active macro cells.

The principle can be summarized as macro-OFF-pico-ON policy. An interesting study is to see how much degradation will be if we restrict the candidate patterns to the pre-selected ones. In Table~\ref{table_gap}, we show results obtained by our joint strategy but allowing only four candidate patterns constructed according to the above criterion. The results are given in percentage of the original performance achieved without any restriction of candidate patterns. As shown, all metrics are very close to the original performance.

\begin{table}[!ht]
\renewcommand{\arraystretch}{1.0}
\caption{Percentage of the original performance if the pre-selection criterion is used for the proposed joint strategy.}
\label{table_gap}
\centering
\begin{tabular}{|c|c|c|c|c|}
  \hline
   & Log-utility  & 5th-\%tile  & Median & Sum rate \\
   \hline
  90 UEs  & 99.72\% & 94.09\% & 94.64\%  & 98.93\% \\
  180 UEs & 99.76\% & 96.25\% & 97.63\% & 99.09\%\\
  300 UEs & 99.77\% & 94.42\% & 96.96\% & 99.95\% \\
  \hline
\end{tabular}
\end{table}

\section{Conclusion}

In this paper, we have studied the joint user association and ICI management in heterogeneous networks. We treated the multi-cell frequency allocation as frequency partitioning among multiple reuse patterns and developed algorithms for joint optimization of user association and reuse pattern selection. We identified that although the number of all possible patterns grows exponentially with the number of cells, most of the patterns are not used as the result of the optimization. The similar observation has been made in TDMA homogeneous network before. However, in order to select the important candidate patterns, a different criterion is needed for the heterogeneous network due to the overlaid deployment. Finally, we have shown that our algorithm achieves significant performance gain compared to the conventional universal reuse scheme.



\end{document}